\documentclass[preprint,12pt]{elsarticle}

\usepackage{graphics}

\biboptions{sort&compress}

\journal{Nucl. Instr. and Meth. A}

\begin{document}

\begin{frontmatter}



\title{A reaction plane detector for PHENIX at RHIC\tnoteref{t1}}
\tnotetext[t1]{For the PHENIX Collaboration.}


\author[UMD]{E. Richardson\corref{cor1}}
\cortext[cor1]{Corresponding author}
\ead{ericr@umd.edu}
\author[RIKJ,RIK]{Y. Akiba}
\author[UMD]{N. Anderson}
\author[UC]{A.A. Bickley}
\author[UT]{T. Chujo}
\author[CU]{B.A. Cole}
\author[UT]{S. Esumi}
\author[BNL]{J.S. Haggerty}
\author[CU]{J. Hanks}
\author[SBUP]{T.K. Hemmick}
\author[UMD]{M. Hutchison}
\author[UT]{Y. Ikeda}
\author[UT]{M. Inaba}
\author[BNL,SBUC]{J. Jia}
\author[BNL]{D. Lynch}
\author[UT]{Y. Miake}
\author[UMD]{A.C. Mignerey}
\author[UT]{T. Niida}
\author[BNL]{E. O'Brien}
\author[BNL]{R. Pak}
\author[UT]{M. Shimomura}
\author[ONL]{P.W. Stankus}
\author[UT]{T. Todoroki}
\author[UT]{K. Watanabe}
\author[SBUC]{R. Wei}
\author[RIK]{W. Xie}
\author[CU]{W.A. Zajc}
\author[ONL]{C. Zhang}

\address[BNL]{Physics Department, Brookhaven National Laboratory, Upton, New York 11973-5000, USA}
\address[UC]{University of Colorado, Boulder, Colorado 80309, USA}
\address[CU]{Columbia University, New York, New York 10027 and Nevis Laboratories, Irvington, New York 10533, USA}
\address[UMD]{University Of Maryland, College Park, Maryland 20742, USA}
\address[ONL]{Oak Ridge National Laboratory, Oak Ridge, Tennessee 37831, USA}
\address[RIKJ]{RIKEN Nishina Center for Accelerator-Based Science, Wako, Saitama 351-0198, JAPAN}
\address[RIK]{RIKEN BNL Research Center, Brookhaven National Laboratory, Upton, New York 11973-5000, USA}
\address[SBUC]{Chemistry Department, Stony Brook University, SUNY, Stony Brook, New York 11794-3400, USA}
\address[SBUP]{Department of Physics and Astronomy, Stony Brook University, SUNY, Stony Brook, New York 11794, USA}
\address[UT]{Institute of Physics, University of Tsukuba, Tsukuba, Ibaraki 305, Japan}

\begin{abstract}
A plastic scintillator paddle detector with embedded fiber light guides and 
photomultiplier tube readout, referred to as the Reaction Plane Detector (RXNP), 
was designed and installed in the PHENIX experiment prior to the 2007 run of the 
Relativistic Heavy Ion Collider (RHIC).  The RXNP's design is optimized to accurately 
measure the reaction plane (RP) angle of heavy-ion collisions, where, for mid-central 
$\sqrt{s_{NN}}$ = 200 GeV Au+Au collisions, it achieved  a $2^{nd}$ harmonic RP 
resolution of $\sim$0.75, which is a factor of $\sim$2 greater than PHENIX's 
previous capabilities.  This improvement was accomplished by 
locating the RXNP in the central region of the PHENIX experiment, where, due to its
large coverage in pseudorapidity ($1.0<|\eta|<2.8$) and $\phi$ (2$\pi$), it is exposed
to the high particle multiplicities needed for an accurate RP
measurement.  To enhance the observed 
signal, a 2-cm Pb converter is located between the nominal collision region and 
the scintillator paddles, allowing neutral particles produced in the heavy-ion 
collisions to contribute to the signal through conversion electrons.  This paper 
discusses the design, operation and performance of the RXNP during the 2007 RHIC run.
\end{abstract} 


\end{frontmatter}



\section{Introduction}\label{sec:intro}

The Reaction Plane Detector (RXNP) was installed in the PHENIX 
experiment~\cite{ref_phenix}, located at Brookhaven National Laboratory's 
Relativistic Heavy Ion Collider (RHIC)~\cite{ref_rhic}, prior to the 2007 RHIC run.
The RXNP is a scintillator paddle detector embedded with optical fiber light 
guides connected to photomultiplier tubes (PMT's), with the design purpose of 
accurately measuring the reaction plane (RP) angle of heavy ion collisions,
which is defined by the impact parameter and beam axis, and whose angle 
($\Psi_{R}$) is determined with respect to a constant arbitrary angle 
from the laboratory coordinate system.  To increase the accuracy of this measurement a 2-cm lead (Pb) converter is located directly in front of the scintillators with respect to the nominal collision region, thereby allowing neutral particles to contribute to the signal through conversion electrons, as well as increasing the overall particle flux through the scintillators and thus increasing energy deposition.  However, due mainly to 
finite particle statistics and detector granularity, it is 
impossible to know $\Psi_{R}$ with absolute certainty, thus its experimental measurement is referred to as the event plane (EP) angle. 

Measuring the EP angle is important in heavy-ion collisions because it allows for the
study of how the collective geometry of the participating nucleons affects the emitted particles'
angular abundance, such as in elliptic flow ($v_{2}$), the path length dependence
of their energy loss through the medium, as well 
as being an important component when comparing heavy-ion and p+p collisions.
Such studies provide a wealth of information about the properties of
the created matter.  For instance, identified particle scaling in $v_2$ measurements has 
shown that the medium exhibits quark degrees of 
freedom~\cite{ref_quark_scaling_phnx, ref_quark_scaling_star}, thereby 
lending strong evidence to the formation of a quark gluon plasma (QGP).  These $v_2$ 
measurements are well described by hydrodynamic models at low transverse 
momentum ($p_{T}$)~\cite{ref_hydro}, suggesting the medium behaves like an ideal
fluid.  Furthermore, $v_2$ studies, along with the geometrical dependence of 
jet suppression~\cite{ref_jet_supp_phnx, ref_jet_supp_star}, 
which also relies on the event-by-event measurement of the EP angle,
were an important basis for the conclusion in the RHIC white papers~\cite{ref_white_phenix,
ref_white_star, ref_white_phobos, ref_white_brahms} 
that a highly dense, strongly interacting medium is formed in RHIC's Au+Au collisions when the center-of-mass energy per nucleon pair ($\sqrt{s_{NN}}$) is 200 GeV.  In addition, many other measurements 
rely on measuring the EP angle, including 
other harmonic flow measurements, CP violation, HBT, and 
$R_{AA}$~\cite{ref_v1, ref_v1_v4, ref_cp_violation, ref_hbt_1, ref_hbt_2, ref_Raa}.

The RXNP is used to measure the EP angle from the azimuthal asymmetry of the produced particle distribution, 
referred to as anisotropic flow, which can be described by the Fourier 
expansion~\cite{ref_rp_method}
\begin{equation}
\frac{d(wN)}{d(\phi - \Psi_R)} = \frac{\langle wN \rangle}{2\pi} \left(1+\sum_{n} 2v_{n}\cos[n(\phi - \Psi_{R} )] \right),
\end{equation}
where $n$ represents the $n$th harmonic of the distribution,
$N$ is the number of particles measured, $\phi$ is the particle
angle, $w$ are weights, and $v_{n}$ is the anisotropy parameter representing the magnitude 
of the flow signal.  From this same anisotropy the EP angle is measured using
\begin{equation}\label{eqn:rp_angle}
\Psi_n = \frac{1}{n}\tan^{-1}\left(\frac{\displaystyle{Y_n=\sum_{i} w_{i}\sin(n\phi_i)}}{\displaystyle{X_n=\sum_{i} w_{i}\cos(n\phi_i)}}\right),
\end{equation}
where $\Psi_{n}$ is the measured EP angle from the $n$th harmonic particle distribution and  $X_n$ and $Y_n$ are the event flow vectors.

Before the installation of the RXNP, PHENIX measured $\Psi_{n}$ using it's Beam Beam
Counter (BBC) detectors~\cite{ref_bbc}, which, for mid-central $\sqrt{s_{NN}}$ = 200 GeV Au+Au 
collisions, had a $2^{nd}$ harmonic EP resolution, defined as $\langle\cos[n(\Psi_n-\Psi_{R})]\rangle$, of $\sim$0.4, where 1.0 would 
denote a perfect resolution.  This resolution proved sufficient for studying 
abundant low $p_{T}$ particles, but was insufficient for making new discoveries 
with photons and rarer probes such as $J/\psi$ and high $p_{T}$ particles. 
Therefore, a better EP resolution was required to increase the physics capabilities of 
a data set by yielding smaller error bars during analysis.  The RXNP was designed 
and built to fulfill this need.

Reported here are the simulations and experimental tests done to optimize the RXNP's design and material compositions.  The final detector geometry is also discussed, along with its online 
performance during the 2007 RHIC run where Au nuclei were collided at 
$\sqrt{s_{NN}}$ = 200 GeV.  Finally, a discussion of the detector's calibrations
and performance is included.

\section{Simulations and Testing}\label{sec:sim_test}
\begin{figure}
\begin{center}
\includegraphics[width=0.7\textwidth]{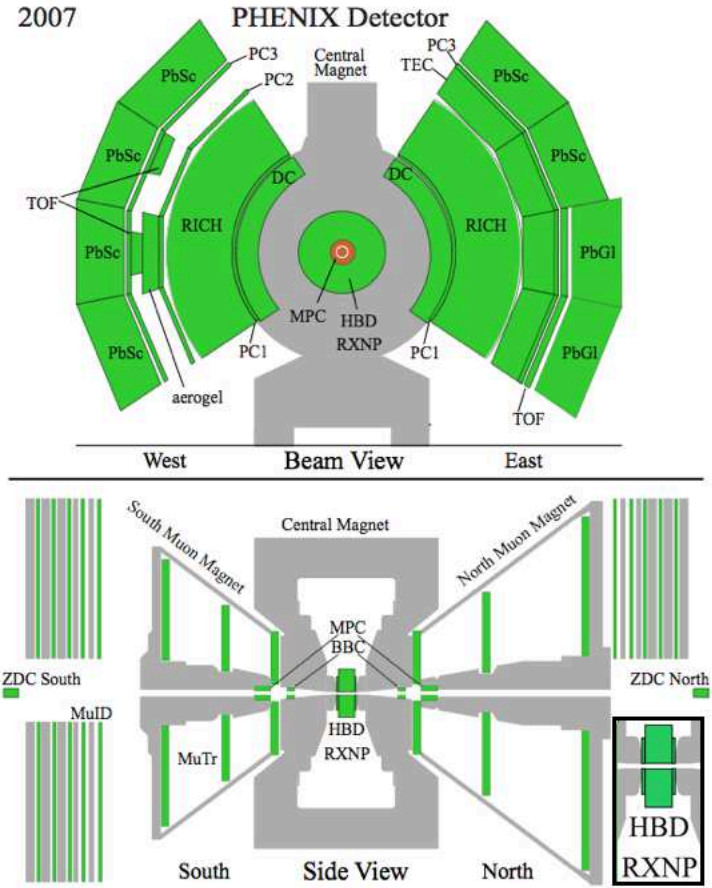}
\end{center}
\caption{\label{sim:phenix_det}
PHENIX detector configuration for the 2007 RHIC run.  As shown in the Side View insert (bottom right), the RXNP is sandwiched between the central magnet and the HBD.}
\end{figure}

The RXNP was designed to optimize the resolution of the $2^{nd}$ harmonic EP measurement, while not interfering with the location and particle acceptance of existing PHENIX sub-systems.  One contributing factor that strongly influences the resolution is the particle multiplicity that is incident on the detector, where a large multiplicity is desirable.  This can be maximized by designing the RXNP to include a large pseudorapidity ($\eta$) coverage, while placing it close to the nominal vertex position.  Within the PHENIX experiment, shown in Fig.~\ref{sim:phenix_det}, this region is largely occupied by existing sub-systems, magnets and support structures.  After considering the space available for a new detector, it was decided that the RXNP would be composed of two mirror image halves of radiating paddles located approximately $\pm$40 cm from the nominal vertex position and attached to the face of the central magnet's copper (Cu) nosecones.

This location provided several design challenges including: (1) there being 
only 7 cm of available space between each nosecone and the soon to be installed
Hadron Blind Detector (HBD)~\cite{ref_hbd} and
(2) the detector had to be able to operate effectively in the high magnetic field
environment of PHENIX's central region, where the field strength can be as high 
as 1 Tesla.  In order to satisfy these requirements 
and answer many outstanding design questions, including
material composition, converter effectiveness, and signal readout,  extensive
studies were performed.

Some of these issues were addressed in GEANT3~\cite{ref_geant3} based simulations using realistic
$v_{2}$ and multiplicity distributions in $\eta$, $p_{T}$ and 
$\phi$ for $\sqrt{s_{NN}}$ = 200 GeV 
Au+Au collisions.  The RXNP was modeled as two scintillating disks surrounding the 
beam pipe and evenly divided into 8 segments in $\phi$, while located at either $\pm$34 or 
$\pm$39 cm from the nominal vertex.  The $\eta$ coverage for these simulations 
varied, but was always between $0.8<|\eta|<2.8$.  Placed immediately in front of the
scintillators was a similarly shaped metal converter whose purpose was to increase the EP resolution through the means described in Sec.~\ref{sec:intro}.

\begin{figure}
\begin{center}
\includegraphics[width=0.8\textwidth]{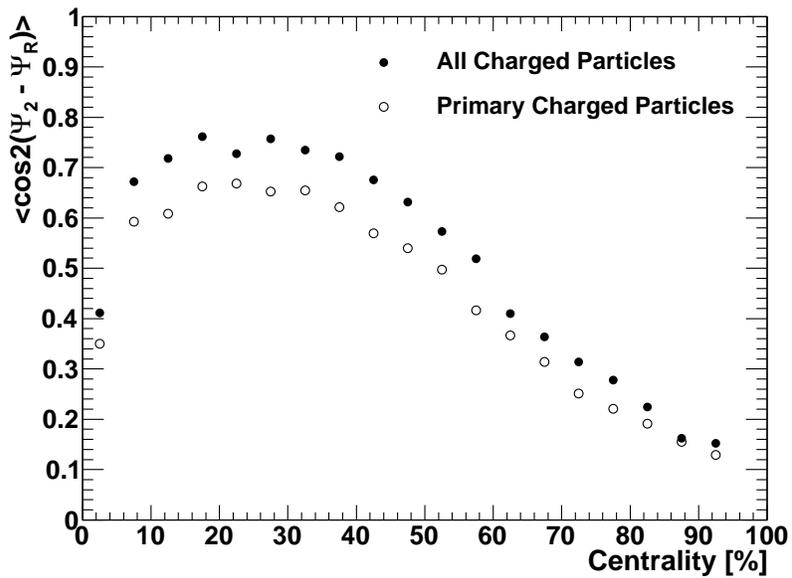}
\end{center}
\caption{\label{sim:rp_res}
Second harmonic EP resolution vs. centrality of the RXNP using a GEANT3 based simulation, where open 
circles pertain to only primary charged particle hits and closed circles are for all 
charged particle hits.
Here the effectiveness of the converter is demonstrated by the secondary charged particles 
increasing the detector's resolution compared to using only 
primary charged  particles.  Also notice the resolution varies as a function of centrality, where it is at a maximum between $20\%$ and $30\%$.  This is caused by a combination of changing event multiplicity and $v_{2}$ signal.}
\end{figure}

\begin{figure}
\begin{center}
\includegraphics[width=0.99\textwidth]{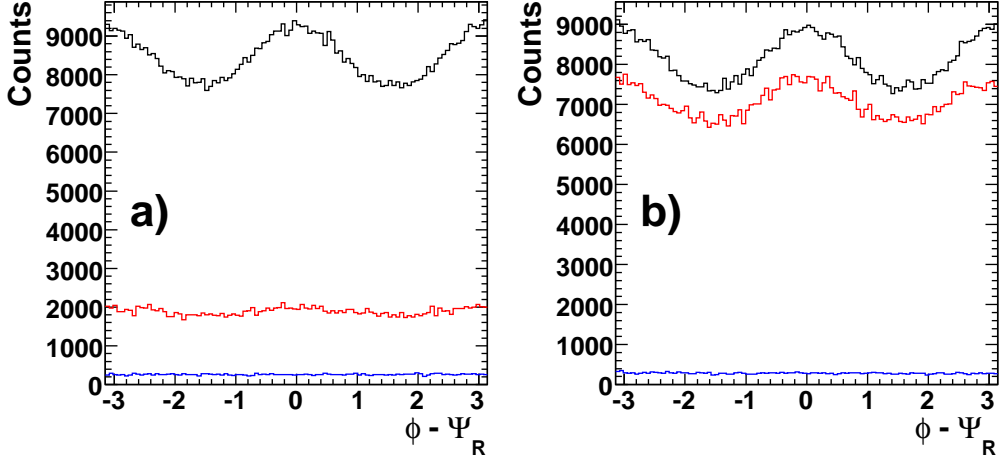}
\end{center}
\caption{\label{sim:particle_dist}
Particle distribution with respect to $\Psi_{R}$ for primary (top, black), 
secondary (middle, red) and background (bottom, blue) particles
without ($a$) and with ($b$) a converter.  Without the converter the secondary particles
have a nearly flat distribution, but with a 0.5 cm brass converter the secondary 
particles exhibit a strong flow signal, thereby reinforcing the flow signal from the
primary particles.}
\end{figure}

The effectiveness of the converter in accomplishing this is
seen in Fig.~\ref{sim:rp_res}, which shows resolution vs. centrality, $i.e.$ the percent of collisions having more geometrical overlap than the current collision.  Here an average of 16\% increase in the $2^{nd}$
harmonic EP resolution is seen when all the charged particles of a collision are used to
measure the EP, compared to using only primary charged particles.
A major reason for this improvement is shown in Fig.~\ref{sim:particle_dist} 
where the primary, secondary and background particle distributions are shown
with respect to $\Psi_{R}$.  With no converter ($a$)
a strong flow signal from primary particles is seen, but this is diluted by the secondary
particles showing a nearly flat distribution.  However, if a 0.5-cm brass converter is 
added, as seen in ($b$), the secondary particles not only increase in number due to the production of conversion electrons, as demonstrated by
a similar simulation in Table~\ref{tbl:q_particles}, but they also
carry a strong flow signal themselves that originates from the distribution of the parent particles, thereby reinforcing the flow signal from the 
primary particles.  The addition of a
converter was also shown to reduce the low energy background.

\begin{table}[hctb]
\caption{Charged particles per segment using different thickness brass converters.}
\label{tbl:q_particles}
\begin{center}
\begin{tabular}{|cccc|}
\hline
Converter Thickness & Primary & Secondary & Total\\
\hline
No Converter & 151 & 78  & 229\\
1.5 cm       & 132 & 340 & 472\\
4 cm         & 103 & 723 & 826\\
\hline
\end{tabular}
\end{center}
\end{table}

\begin{figure}
\begin{center}
\includegraphics[width=0.99\textwidth]{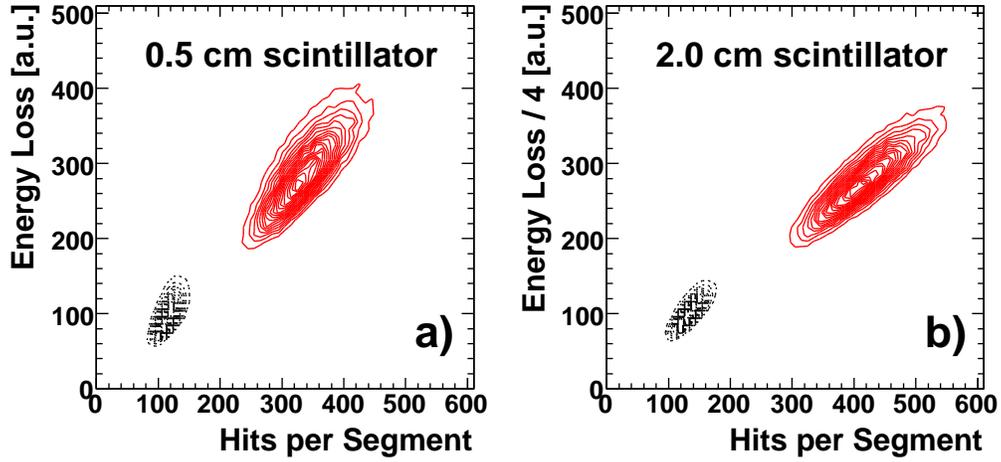}
\end{center}
\caption{\label{sim:scint_energy}
Number of particle hits per scintillator segment vs. energy loss.
The scintillator thickness is 0.5 and 2.0 cm for ($a$) and ($b$), respectively.
The dotted contour lines (black) use no converter, while
the solid lines (red) use a 2.0-cm brass converter.  Notice the energy loss 
in ($b$) is divided by 4.  Both with and
without the converter, the correlation is better with the thicker scintillator.}
\end{figure}

\begin{figure}
\begin{center}
\includegraphics[width=0.70\textwidth]{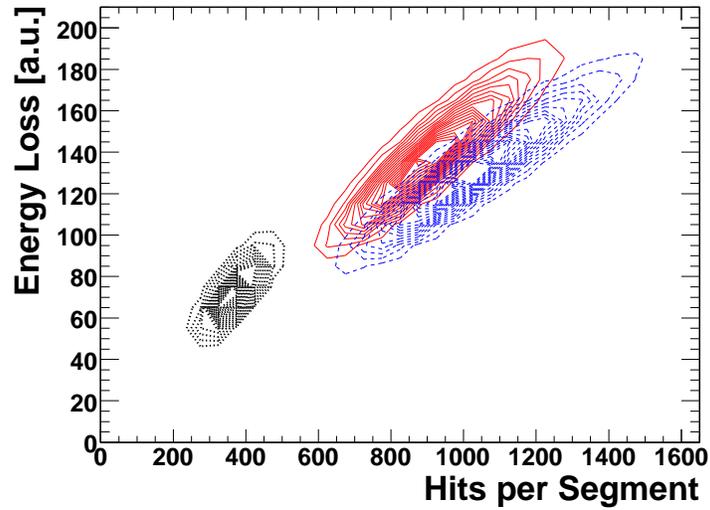}
\end{center}
\caption{\label{sim:energy_dep_conv}
Particle hits per scintillator segment vs. energy loss using 2.0-cm brass (dotted, black),
Pb (solid, red) and W (dashed, blue) converters.  Tungsten provides the best hit
to energy loss correlation, but only marginally so over Pb, which is significantly 
less expensive.  In all cases a 2-cm scintillator was used.}
\end{figure}

Increasing converter thickness was also shown to increase the $2^{nd}$
harmonic EP resolution, as seen in Table~\ref{tbl:converter_res}, while
increasing scintillator thickness was shown to give a better correlation between the number
of particle hits and energy deposition, as depicted in Fig.~\ref{sim:scint_energy}.  
To optimize the converter and scintillator thickness within the constraints of the limited space available between the nosecone and HBD, 
it was decided to restrict the thickness of the converter and scintillator to 2 cm each. 
Therefore, 2-cm thick brass, Pb and tungsten (W) converters were simulated 
with a 2-cm thick scintillator to compare energy deposition, as shown in 
Fig.~\ref{sim:energy_dep_conv}, and their effect on the $2^{nd}$
harmonic EP resolution for mid-central collisions, which were found to be
0.70, 0.74 and 0.76, respectively.  In both cases W performed best, but only 
marginally so compared to Pb, which is significantly less expensive.
Therefore, Pb was chosen as the converter material.

\begin{table}[hctb]
\caption{Simulated $2^{nd}$ harmonic EP resolution for mid-central events using
different thickness converters and a 2-cm thick scintillator.}
\label{tbl:converter_res}
\begin{center}
\begin{tabular}{|cccccc|}
\hline
Converter Thickness (cm) & 0.0  & 1.0  & 2.0  & 4.0  & 8.0\\
\hline
Brass                    & 0.53 & 0.65 & 0.71 & 0.76 & 0.80\\
Pb                       & 0.53 & 0.73 & 0.75 & 0.75 & 0.68\\
\hline
\end{tabular}
\end{center}
\end{table}

\begin{figure}
\begin{center}
\includegraphics[width=0.60\textwidth]{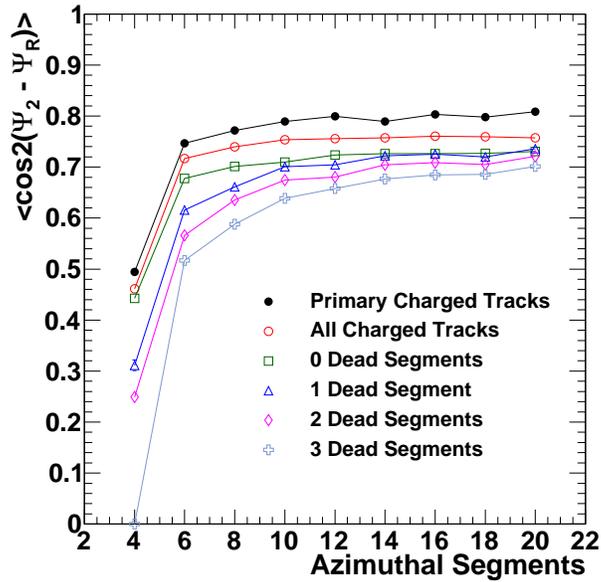}
\end{center}
\caption{\label{sim:scint_seg}
Dependence of the $2^{nd}$ harmonic EP resolution of mid-central collisions on the azimuthal segmentation of the RXNP scintillator.  The closed circles use only the number of 
primary charged particles in each segment as the weight,
with each particle having a weight of 1.  The resolution for the open circles is similarly
calculated, but uses all charged particles (primary, secondary and background).  The square, 
triangle, diamond and cross data points all calculate the resolution using the energy 
deposition into the scintillators from charged particles as the weighting factor, while at the same time 
having 0, 1, 2 or 3 dead scintillator segments, respectively.  In all cases a 2-cm 
scintillator and 2-cm brass converter were used.}
\end{figure}

\begin{figure}
\begin{center}
\includegraphics[width=0.60\textwidth]{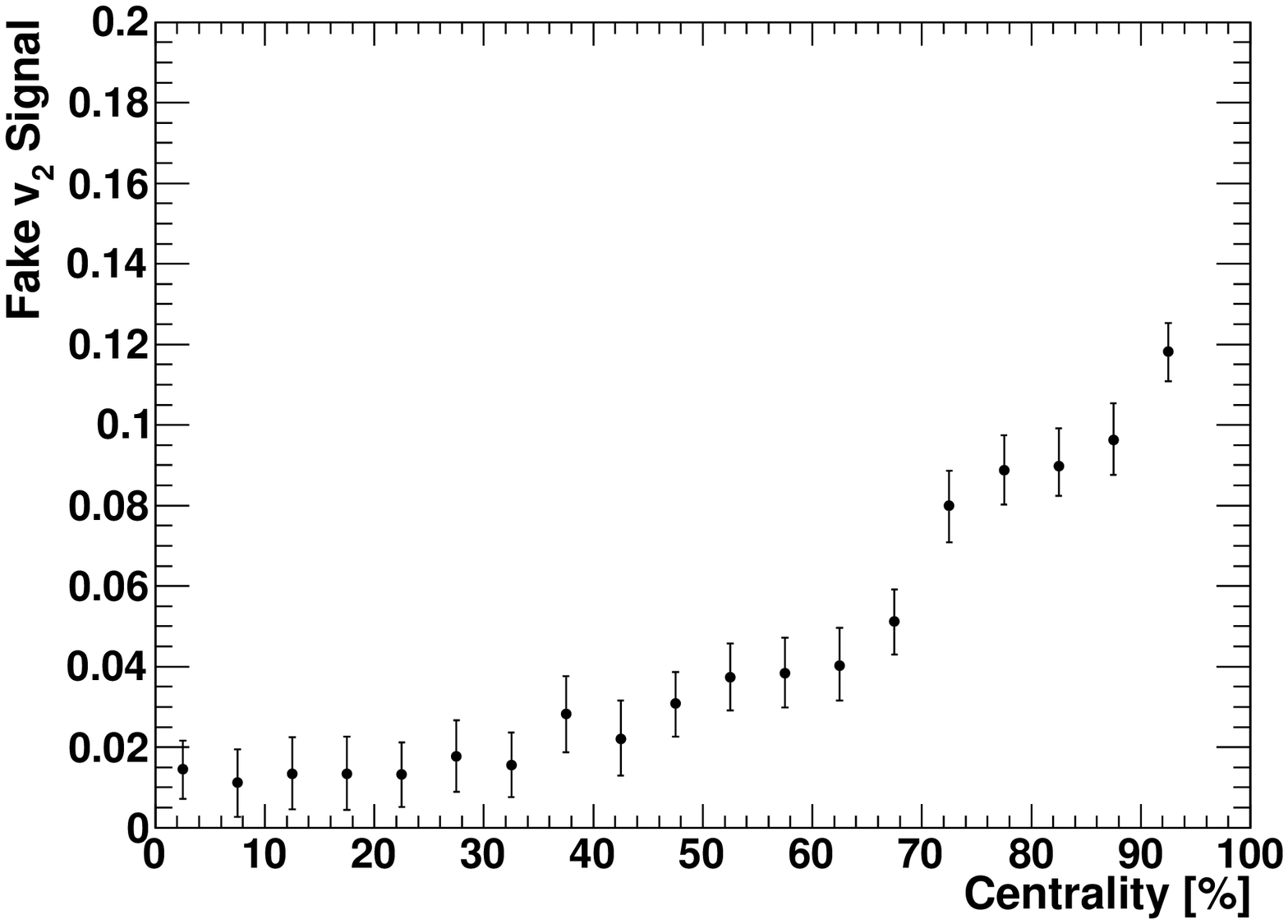}
\end{center}
\caption{\label{sim:fake_v2_cent}
Simulated fake $v_{2}$ signal from jets with respect to centrality in the PHENIX 
central arm spectrometers ($|\eta|<0.35$) when using the EP calculated from the 
RXNP and requiring each event's leading hadron 
to have a $p_{T}>6$ GeV/$c$.  1 $\sigma$ statistical error bars are also shown.}
\end{figure}

The optimum radiator $\phi$ segmentation for the $2^{nd}$ harmonic EP resolution 
was also studied.  Figure~\ref{sim:scint_seg} shows that the resolution improves
only marginally above 8 segments; however, to protect against failing segments it was 
decided to partition the detector into 12 paddles in $\phi$.  
The RXNP was further divided into two radial sections covering $1.0<|\eta|<1.5$ (outer ring) and $1.5<|\eta|<2.8$ (inner ring) as a result of a series of simulations incorporating the PHENIX central arm spectrometers ($|\eta|<0.35$) that showed a centrality and $\eta$ dependent fake $v_{2}$ signal from jets when using the EP of the RXNP.  These studies showed that the largest bias occurs in peripheral events, shown 
in Fig.~\ref{sim:fake_v2_cent}, and the closer in proximity the RXNP's $\eta$ coverage is to the central arms.  This bias can result from back-to-back di-jet correlations, as well as correlated particle production from the near-side jet cone, which can have a size of 0.7 units in $\eta$.  

Therefore, to minimize the impact of this bias on a wide range of physics analyses, the RXNP was divided into two $\eta$ sections resulting in the inner ring section experiencing only a minimal bias effect.  This added 
flexibility was proven effective in~\cite{ref_ppg110}, where only the inner 
ring was used in measuring the $v_{2}$ of high $p_{T}$ $\pi^{0}$'s,
and in~\cite{ref_ppg098}, which used both $\eta$ sections in examining the 
$\eta$ dependence of non-flow effects.

Avalanche photodiodes (APD's) and PMT's 
were considered for signal readout.  APD's have several advantages over PMT's,
including being unaffected by magnetic fields, significantly cheaper 
and allowing for a simpler design by jettisoning the 
need for light guides.  However, after performing calculations and
bench tests, it was confirmed that the signal to noise ratio from APD's was insufficient for this application.  Therefore, it was decided to pursue a PMT design.

To confirm that PMT's could function well in PHENIX's strong magnetic field
and to investigate an optimal radial position and angular orientation with
respect to the field lines, a series of magnetic field tests were performed.  
One of these tests involved constructing and positioning a test stand in PHENIX's
central region that simultaneously held 2 sets of 4 PMT's, separated by $120^\circ$,
at radial positions where their gains were predetermined to be best: 80, 90, 110 and 130 cm.  
Hamamatsu R5924 fine mesh PMT's were used, which are specifically designed to
operate in a high magnetic field environment.  Using an LED pulser, each tube's gain was recorded
at zero and full field for PHENIX's ``++'' and ``+-'' field configurations.  In addition, because the region in which the RXNP was to be installed is near the magnetic field return, the field lines exhibit non-uniform behavior that is not well known.  Moreover, with PMT performance in a magnetic field known to be sensitive to the alignment of the tube with respect to the field lines, two configurations were tested for each field setting: PMT's aligned parallel and at $30^{\circ}$ to the beam pipe. 
The results of this test are shown in Table~\ref{tbl:mag_field_test}, where the 
numbers are averaged from the two PMT's located at each distance.  
Based largely on these results and the fact that the magnetic field would be 
in the ``+-'' configuration for the 2007 run, it was decided the PMT's would be placed 
parallel to the beam pipe at a radial distance of 130 cm.  Here they would
experience a magnetic field strength of $\sim$0.66T ($\sim$0.61T) for the ``+-'' 
(``++'') field configurations.

\begin{table}[hctb]
\caption{Fraction of PMT gain observed at full field compared to zero field.}
\label{tbl:mag_field_test}
\begin{center}
\begin{tabular}{|ccccc|}
\hline
Distance (cm) & ``++'' $30^{\circ}$ & ``+-'' $30^{\circ}$  & ``++'' Parallel & ``+-'' Parallel\\
\hline
80  & 0.61 & 0.00 & 0.76 & 0.56\\
90  & 0.39 & 0.33 & 0.78 & 0.23\\
110 & 0.50 & 0.32 & 0.96 & 0.54\\
130 & 0.44 & 0.35 & 0.85 & 0.66\\
\hline
\end{tabular}
\end{center}
\end{table}

With much of the design solidified, the remaining element to be determined was the radiator and light guide configuration.  This was resolved through beam tests performed at KEK-PS in Japan using a momentum selected charged 
particle beam.  The purpose of these tests were to examine different radiator 
(scintillator vs. Cherenkov)
and light guide (solid vs. embedded fiber) combinations.  These tests showed
that, although the solid light guide resulted in more light collection, it also had a signal
size that was dependent on the incident particle's position in the radiator, as 
seen before in~\cite{ref_phobos_paddles}, which would
likely worsen the EP resolution.  This effect was significantly less pronounced using the
embedded fibers.  This, combined with the fibers reasonable light collection and
flexibility in positioning and grouping PMT's, led to their selection.
The Cherenkov radiator was 
eliminated because it yielded too small a signal when used with embedded fibers, while
the scintillator signal was reasonable.  Thus scintillator radiators with embedded fiber
light guides were chosen for the RXNP.

\section{Design and Geometry}
\begin{figure}
\begin{center}
\includegraphics[width=0.70\textwidth]{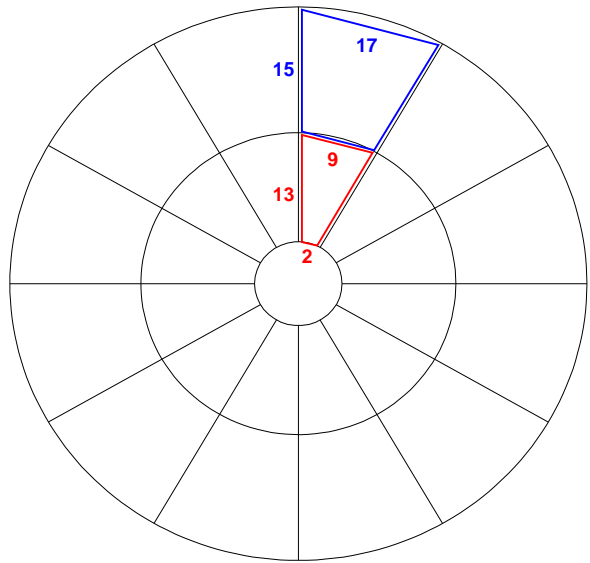}
\end{center}
\caption{\label{design:scint_dimensions}
Schematic diagram illustrating the arrangement of the inner (red) and outer (blue)
scintillator rings.  The length of each scintillator side is shown in cm's.}
\end{figure}

The RXNP is composed of two sets of 24 scintillators, a north (N) and a south (S),
located $\pm39$ cm from the nominal vertex position with the S arm being located 
in the negative direction.  The scintillators are arranged perpendicular 
to and surround a 10-cm diameter beam pipe 
in 2 concentric rings (inner, outer), with each ring having 2$\pi$ coverage and 12 equally sized 
segments in $\phi$.  All scintillators are trapezoidal in shape, 2-cm thick,
made of EJ-200 material from Eljent Technology (equivalent to BC408) and individually
wrapped with an inner layer of aluminized mylar sheeting for light reflection and an outer
layer of black plastic for light tightness.  A schematic diagram showing the arrangement  
of the scintillators and their sizes is shown in Fig.~\ref{design:scint_dimensions}, 
with the edges of the inner ring positioned at radial distances of 5 and 18 cm from 
the ion beam covering $1.5<|\eta|<2.8$.  Uninterrupted coverage continues outward with 
the outer ring to 33 cm or $|\eta|=1.0$.  The length of the inner and outer edges 
of the inner scintillators are 2 and 9 cm, respectively, with the outer edge of the
outer scintillators being 17 cm.    

\begin{figure}
\begin{center}
\includegraphics[width=0.70\textwidth]{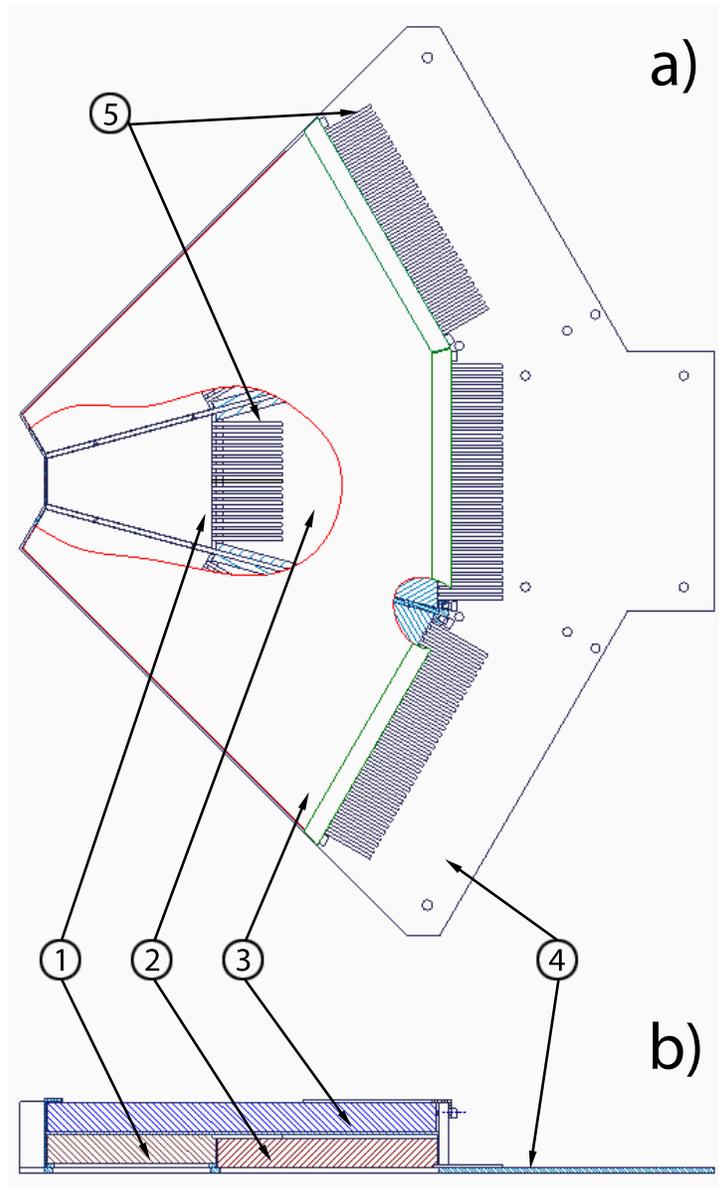}
\end{center}
\caption{\label{design:tray_assembly}
($a$) shows a top view of an assembled tray with a cutout of the Pb 
converter ($3$) showing the inner ($1$) and outer ($2$) scintillators underneath
in their compartments with their optical
fibers ($5$) emerging to their rear.  ($4$) identifies the aluminum tray.
($b$) shows a side view of 
an assembled tray with the Pb converter above the scintillators.  
Notice the plastic spacer directly underneath the 
inner scintillator, offsetting it from the outer scintillator, allowing 
its fiber light guides to exit the tray.}
\end{figure}

Each 24 scintillator set is housed in 4 identical aluminum structures each 
consisting of a tray covering $90^\circ$ in $\phi$ with a support arm extending 
radially $\sim$80 cm.  As shown in Fig.~\ref{design:tray_assembly} ($a$),
each tray contains three compartments for an inner and 
outer scintillator with each scintillator having wavelength shifting fiber light guides 
embedded into its surface every 0.5 cm and running its entire length.  To allow the 
inner scintillator's fibers to run radially out the back of the tray in 
tandem with the outer scintillator's fibers, an offset between the two
is created by placing a 2-mm plastic spacer underneath the inner scintillator,
as shown in Fig.~\ref{design:tray_assembly} ($b$).
On top of the scintillators sits another 2-mm spacer for protection, followed 
by a 2-cm thick converter composed of 98\% Pb doped with 2\% antimony to increase hardness.

As the fibers exit each scintillator they are individually sheathed in black plastic 
tubing for light tightness, bundled together inside flexible plastic tubing for 
stability and protected by an Al cover.  They then run the length of the support arm 
fastened by plastic ties and protected by two more Al covers.
At the end of the arm each scintilltor's fibers are unsheathed from their individual
tubing, yet still encased in the larger flexible tubing, bundled together into a 
plastic end cap and attached to a ``cookie'' covering the face
of a PMT.  Here the cookie guides the light from the fibers into the PMT 
using a $45^\circ$ reflective surface to bend the light a total of 
$90^\circ$.  Along with black tape, a custom built Delrin cap is used to fasten 
the cookie onto the PMT and make the connection light tight.
Each PMT is then fastened to the end of the arm and 
positioned parallel to the beam pipe, giving each quadrant a length of $\sim$124 cm.

\begin{figure}
\begin{center}
\includegraphics[width=0.50\textwidth]{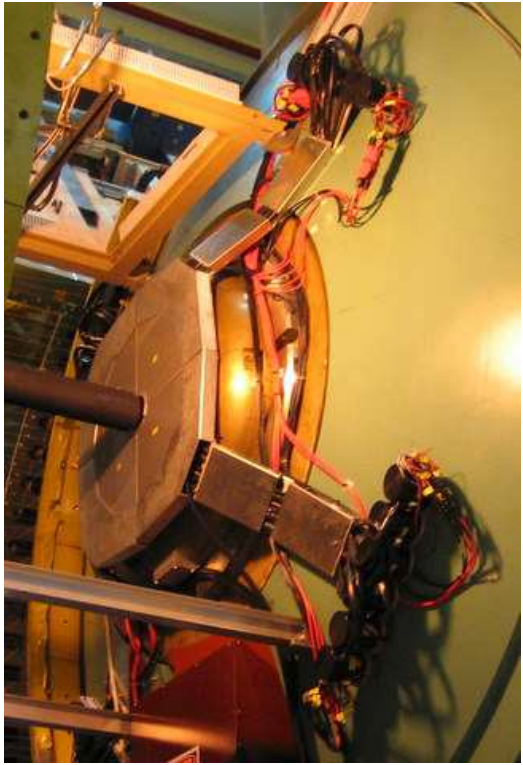}
\end{center}
\caption{\label{design:RXNP_installed}
Picture of the RXNP's north half installed on the Cu nosecone of PHENIX's central 
magnet prior to the installation of the HBD.}
\end{figure}

Hamamatsu R5543 3 inch fine mesh PMT's are 
used to measure the signal.  Although not the same type of PMT used in the original
magnetic field tests discussed in Sec.~\ref{sec:sim_test}, these PMT's are also 
designed to operate in a high magnetic field with similar behavior expected. 
Moreover, these PMT's did undergo their own magnetic 
field testing as mentioned in Sec.~\ref{sec:online}.  Once assembled 
each quadrant was fastened to the nosecone, giving the tray portion of the 
assembly a total thickness of $\sim$5 cm. A picture of the RXNP's north half
after installation is shown in Fig.~\ref{design:RXNP_installed}.

\section{Online Performance and Calibrations}\label{sec:online}
\begin{figure}
\begin{center}
\includegraphics[width=0.60\textwidth]{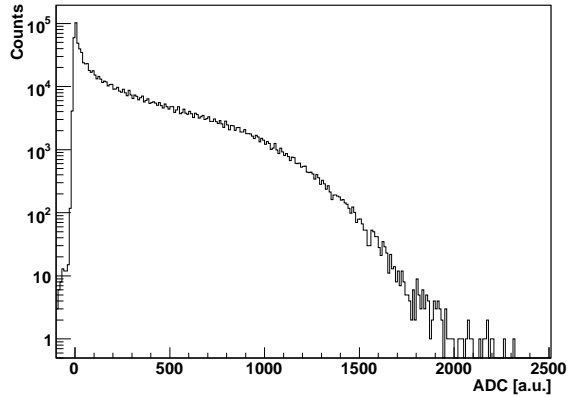}
\end{center}
\caption{\label{online:adc_spectra}
Typical raw ADC spectrum of a PMT from min-bias data.  The distribution results from frequent peripheral (low multiplicity) and rare central (high multiplicity) interactions, which are proportional to the geometrical cross-section.}
\end{figure}

Prior to final assembly and installation, each PMT was tested for noise, 
signal linearity and signal strength in and outside a magnetic field.
After installation one PMT in the south arm became disabled prior to the 2007 RHIC run, 
decreasing the detector acceptance to $\sim$98\% of design, which 
remained throughout the run.  Figure~\ref{online:adc_spectra} shows a typical 
ADC spectrum from a PMT during min-bias data taking, where, at the maximum edge of the distribution, several hundred charged particles, including conversion electrons, are measured in the scintillator segment.  The dynamic range of the ADC is 12 bits or 4096 quantized units, of which $\sim$2000 are typically used.  

\begin{figure}
\begin{center}
\includegraphics[width=0.60\textwidth]{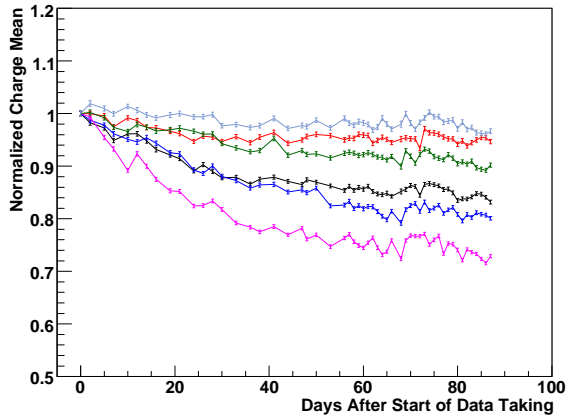}
\end{center}
\caption{\label{online:pmt_deg}
Normalized mean gain values of 6 representative PMT's taken periodically 
during the 2007 RHIC run.  Each PMT is normalized to its own charge mean 
at the start of data taking.  1 $\sigma$ statistical error bars 
are also shown.}
\end{figure}

The gain of the tubes was monitored throughout the run and was found
to decrease over time, as shown in Fig.~\ref{online:pmt_deg}, with no 
pattern seen regarding scintillator location.  During the 2008 and 09 
RHIC runs similar voltages were used while colliding species of 
p+p and d+Au at the same energy, along with p+p at $\sqrt{s_{NN}}$ = 500 GeV.
The p+p data showed no signal degradation, while for d+Au some degradation 
was seen in both RXNP arms, although not as severe as with Au+Au.  For the 2010 
RHIC run, where Au+Au was again collided at $\sqrt{s_{NN}}$ = 200 GeV, 
the RXNP high voltage was increased to counter the loss in gain observed in 2007.  The result was a comparable, although not quite as severe degradation to that of 2007.
The cause of the gain loss remains unexplained, but appears related to the 
use of heavy-ions at higher energies, since no degradation was seen during the 
2010 lower energy ($\le \sqrt{s_{NN}}$ = 62.4 GeV) Au+Au running period.

The observed signal degradation has a negligible effect on the RXNP's 
$2^{nd}$ harmonic EP resolution, except in the 
most peripheral events where the change was less than 5\%.  The effect of the 
removal of the west half of the HBD $\sim$$1/3^{rd}$ 
through the run was also examined and found to be negligible for central events,
but did increase the resolution of peripheral events by up to 10\%.

\begin{figure}
\begin{center}
\includegraphics[width=0.60\textwidth]{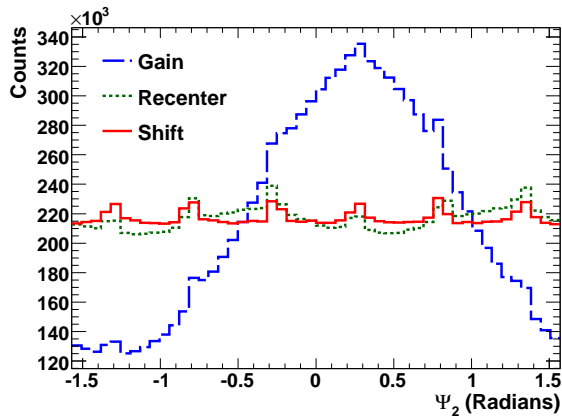}
\end{center}
\caption{\label{online:calib_steps}
Second harmonic EP distribution of min-bias events from a typical run after each 
calibration step.  The dashed line (blue) is after the gain correction, dotted line 
(green) is after recentering and the solid line (red) is after angular shifting.}
\end{figure}

After pedestal subtraction three calibrations were performed on a run by 
run basis to flatten the EP angle distribution, which, by definition, 
should be flat since the RP from the two colliding nuclei has an equal 
probability of occurring at any angle.  However, due to effects such as the
broken PMT, unequal PMT gains, and beam offset,
the EP angular distribution from the raw data was not flat. 
To correct this, each PMT's gain was first calibrated to have the same mean 
ADC value.  The second step recentered the $X_{n}$ and $Y_{n}$ event
flow vector distributions to zero along with adjusting their width to unity 
by~\cite{ref_rp_method}

\begin{equation}
X_n^{corr} = \frac{X_n - \langle X_n \rangle}{\sigma_{X_n}},
\end{equation}

\begin{equation}
Y_n^{corr} = \frac{Y_n - \langle Y_n \rangle}{\sigma_{Y_n}},
\end{equation}
where $X_n^{corr}$ and $Y_n^{corr}$ are the corrected flow vectors and 
$\sigma_{X_n}$ and $\sigma_{Y_n}$ are the uncorrected distribution widths.  The third 
step fits a Fourier expansion to the modified distribution and performs an event-by-event
shifting of the angles using~\cite{ref_shift}

\begin{equation}
\Psi_n^{corr} = \Psi_n + \frac{1}{n}\displaystyle{\sum_i \frac{2}{i} \left[- \langle\sin(in\Psi_n)\rangle \cos(in\Psi_n) + \langle\cos(in\Psi_n)\rangle \sin(in\Psi_n) \right]},
\end{equation}
where $\Psi_n^{corr}$ is the corrected and final EP angle.  The result from each of 
these steps for the $2^{nd}$ harmonic EP is shown in Fig.~\ref{online:calib_steps}.  Notice 
the final distribution has six remaining ``spikes''.  These spikes result from the flattening
of the distribution to the $5^{th}$ harmonic in the final calibration step, along with the 
combination of finite detector granularity and low multiplicity peripheral events.  For 
these events it is possible for only one paddle to be hit causing its angular center to be assigned as that event's EP angle.  This causes an 
overrepresentation of that angle in the EP distribution that is not eliminated
by the calibrations.  However, these spikes do disappear for event centralities $<70\%$.

\section{EP Resolution}
The RXNP can measure the EP angle from nine different detector segment combinations;
N+S, N+S inner ring, N+S outer ring, N (S), N (S) inner ring and N (S) outer ring.  
Two methods can be used to determine the resolution of the different detector segments, 
which are explained in detail in Ref.~\cite{ref_rp_method}, but will be briefly outlined 
here.  The first method uses the EP angle from two equal multiplicity 
subevents, ($a$) and ($b$),  where the resolution of each subevent is
\begin{equation}
\langle\cos[km(\Psi_m^{a} - \Psi_R)]\rangle = \sqrt{\langle\cos[km(\Psi_m^{a} - \Psi_m^{b})]\rangle},
\end{equation}
with $m$ denoting the harmonic of the measured EP being used and $k$ 
is a multiplier such that $n=km$.
The resolution of the combined subevents can be determined using 
\begin{equation}~\label{eq:two_sub}
\langle\cos[km(\Psi_m - \Psi_R)]\rangle = \frac{\sqrt{\pi}}{2\sqrt{2}} \chi_m \exp( - \chi_m^{2}/4)\times[ I_{(k-1)/2} (\chi_m^{2}/4)+I_{(k+1)/2} (\chi_m^{2}/4)],
\end{equation}
where $\chi_m = v_m \sqrt{2N}$ with $v_m$ being the measured flow signal
and $I_{(k-1)}$ and $I_{(k+1)}$ are modified Bessel functions.
The second method uses three subevents where the multiplicity of each is not
required to be equal.  The resolution for this method is calculated by
\begin{equation}\label{eq:three_sub}
\langle\cos[km(\Psi_m^{a}-\Psi_{R})]\rangle = \sqrt{\frac{\langle\cos[km(\Psi_m^{a}-\Psi_m^{b})]\rangle \; \langle\cos[km(\Psi_m^{a}-\Psi_m^{c})]\rangle}{\langle\cos[km(\Psi_m^{b}-\Psi_m^{c})]\rangle}}.
\end{equation}

\begin{figure}
\begin{center}
\includegraphics[width=0.49\textwidth]{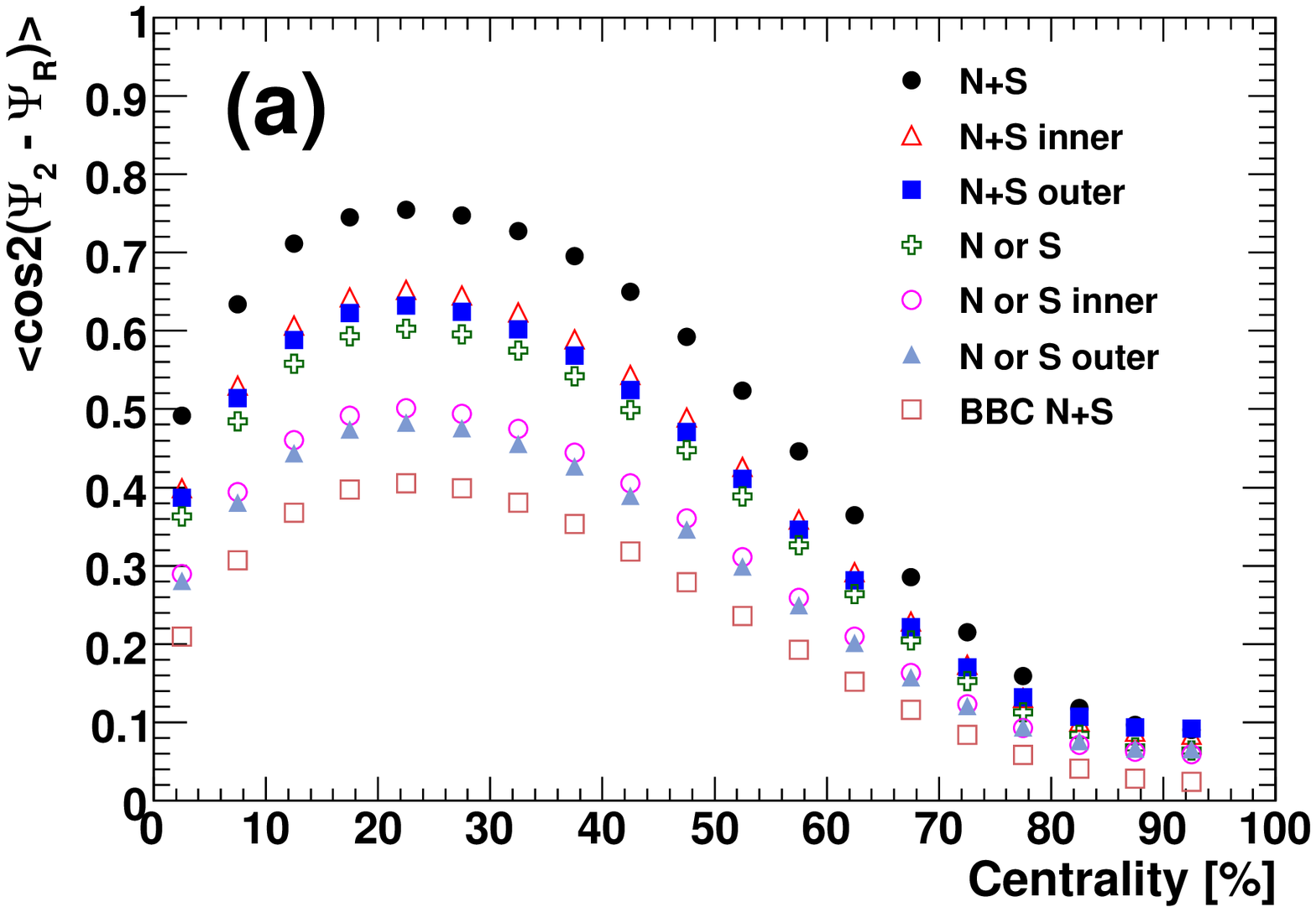}
\includegraphics[width=0.49\textwidth]{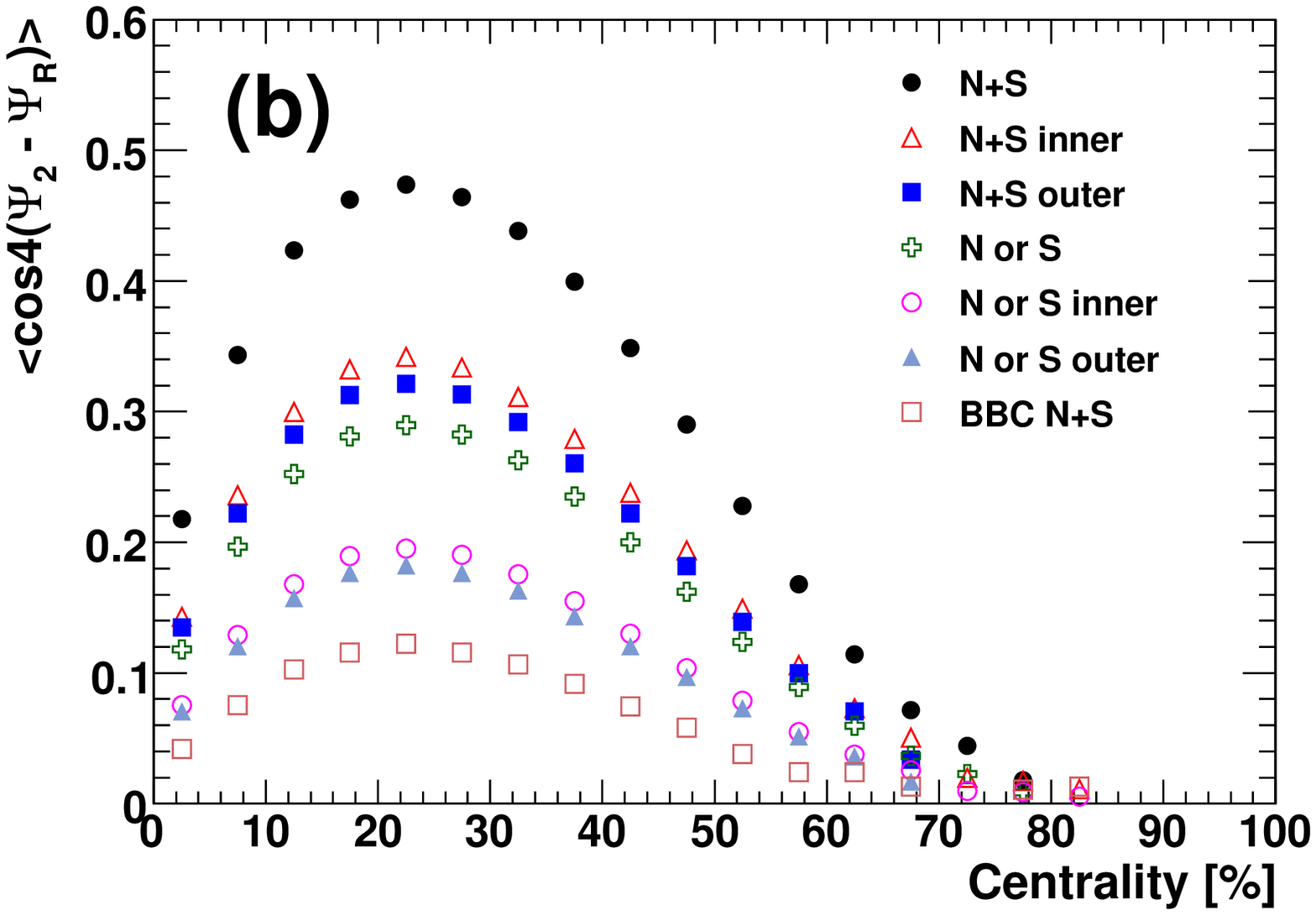}
\end{center}
\caption{\label{offline:rxnp_res}
($a$) shows the $2^{nd}$ harmonic EP resolution for different RXNP detector 
segments using min-bias events, while ($b$) is the $4^{th}$ harmonic.  For reference, the 
corresponding BBC resolution is also shown.  Notice ($a$) and ($b$) have different size y-axis.}
\end{figure}

Using the first method with $m=2$, the $2^{nd}$ and $4^{th}$ harmonic EP resolutions for the 
different segments of the RXNP are shown vs. centrality using min-bias events in 
Fig.~\ref{offline:rxnp_res}.
In this figure one can see a factor of $\sim$2 increase in the $2^{nd}$ harmonic 
EP resolution when using the full RXNP detector compared to the BBC, which is what was 
expected from simulations (see Fig.~\ref{sim:rp_res}).  This factor is even greater 
($\sim$4x)
when using the $4^{th}$ harmonic plane.  The $1^{st}$ harmonic resolution was also 
examined, but showed inconsistencies, making it unreliable.  
These inconsistencies included significantly overestimating the EP resolution for central events
when using the first method.  When using the second method the resolution varied significantly depending upon which of PHENIX's other sub-systems were used in measuring the $\Psi_1^{b}$
and $\Psi_1^{c}$ angles.  Also, when $\Psi_1^{b}$
and $\Psi_1^{c}$ are from each RXNP arm, the resulting resolution of the third
detector's EP angle ($\Psi_1^{a}$) was significantly underestimated.  Possible causes
of this behavior include momentum conservation between the two RXNP arms and the small $1^{st}$ harmonic 
anisotropic flow signal within the RXNP's $\eta$ coverage~\cite{ref_phobos_v1_eta}.

\begin{figure}
\begin{center}
\includegraphics[width=0.60\textwidth]{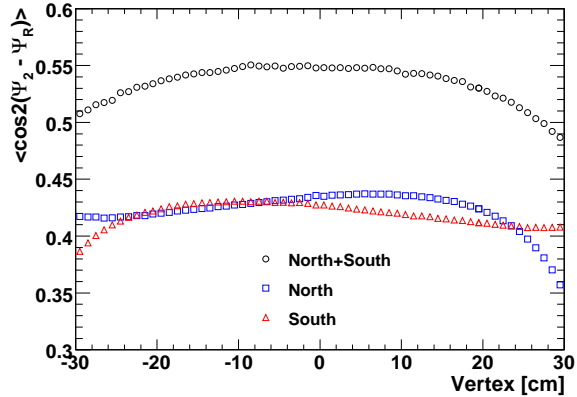}
\end{center}
\caption{\label{offline:rxnp_res_vert}
Second harmonic EP resolution vs. vertex for the south arm (red triangles), north arm
(blue squares) and combined arms (black circles) for min-bias events within the PHENIX vertex acceptance 
of $\pm$30 cm.  The resolution is calculated using the second method (Eq.~\ref{eq:three_sub}) 
where $\Psi^{b}_{2}$ and $\Psi^{c}_{2}$ are from the BBC north and BBC south when 
calculating the combined resolution and RXNP opposite arm and BBC N+S when calculating
the individual arm resolutions.  
In all cases $m=2$.}
\end{figure}

Using the second method (Eq.~\ref{eq:three_sub}), where again $m=2$, the RXNP's 
$2^{nd}$ harmonic EP resolution is shown vs. vertex in Fig.~\ref{offline:rxnp_res_vert}.
This figure shows that each detectors' resolution varies by $<4\%$ within 
$\pm$20 cm of the nominal vertex.  However, at the edge of PHENIX's vertex acceptance 
of $\pm$30 cm the resolution varies by up to 10\% for the south arm, 18\% for the north 
and 11\% when combined.  The effect doubles for the $4^{th}$ harmonic.
For each arm the maximum resolution is achieved at $\sim$7 cm closer to the arm than
the nominal vertex or $\sim$32 cm from each arm. 

The decrease in resolution away from this position is the result of a complex
interplay between a number of factors.
The slow rise in resolution as the vertex approaches a detector arm can largely be explained
by the $\eta$ dependence of $v_{2}$.  As shown in Ref.~\cite{ref_phobos_v2_eta}, the $v_{2}$, 
or $2^{nd}$ harmonic azimuthal particle asymmetry used here to measure the EP, increases as $\eta$ approaches 0, 
resulting in a more accurate EP measurement the closer the event vertex is to an arm 
of the RXNP.  However, this effect is eventually offset due to the decreasing size
of the detector acceptance (smaller $\eta$ coverage) leading to a smaller multiplicity 
of particles used in the EP determination, resulting in a less accurate EP and lower resolution.  

\begin{figure}
\begin{center}
\includegraphics[width=0.6\textwidth]{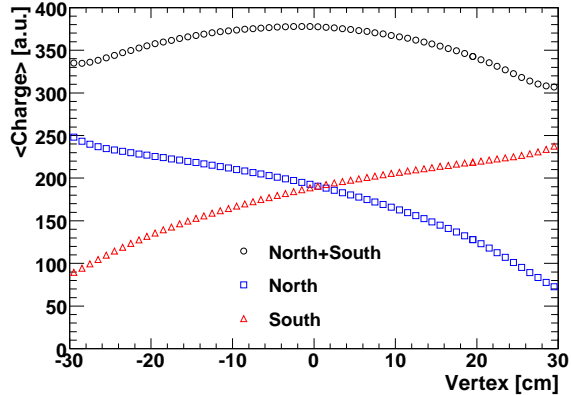}
\end{center}
\caption{\label{offline:charge_vs_vert}
Mean charge deposited into the RXNP south arm (red triangles), north arm (blue squares) 
and combined arms (black circles) as a function of collision vertex position
for min-bias events.  In the case of the individual arms, the charge deposited
increases as the event vertex occurs further away.}
\end{figure}

The change in the observed multiplicity in the RXNP with respect to event vertex is
roughly proportional to the change in energy deposition, as seen
in Fig.~\ref{offline:charge_vs_vert}.
However, it is important to note the closer the event vertex is to an arm, the longer the
average path length the particles have passing through the scintillators, which results in
more energy being deposited on a per particle basis.  This effect acts counter to the
loss in energy deposition due to the lower detector multiplicity, making the energy
deposition only a guide to how the particle multiplicity changes with event vertex. 

\section{Conclusion}
The PHENIX Collaboration successfully designed, installed and commissioned
the RXNP before the 2007 RHIC run to more accurately measure the 
RP angle of colliding Au nuclei.  This detector performed as expected from simulations
by increasing PHENIX's $2^{nd}$ harmonic EP resolution
a factor of $\sim$2 from the previously used BBC detector.  
Combined with the recently completed 2010 RHIC run, where almost 2x the number
of events were collected than in 2007,
this higher resolution detector expands PHENIX's capabilities and allows for
analysis of ever rarer particles and more accurate measurements, which has 
already been demonstrated in several recent 
articles~\cite{ref_ppg110, ref_ppg098, ref_ppg106, ref_ppg116}.  The measurements made possible by the improved EP resolution provided by the RXNP further expand the scientific community's understanding of the QGP.

\section{Acknowledgments}
We would like to thank the KEK E325 experiment for lending the PMT's used
in the RXNP and the PHENIX DAQ team, especially M. L. Purschke and C. Pinkenburg.
We also thank A. Franz and J. La Bounty for their help in the magnetic field testing,
along with M. Lenz for his many contributions.  For the use of 
Fig.~\ref{design:tray_assembly} we thank R. Ruggiero and for 
their contributions we also acknowledge S. P. Stoll, K. Okada and S. Boose.





\bibliographystyle{elsarticle-num}
\bibliography{rxnp_nim}







\end{document}